# Ant Nest Detection Using Underground P-Band TomoSAR

Gian Oré, Alexandre Santos, Daniele Ukan, Ronald Zanetti, Mariane Camargo, Luciano P. Oliveira, Guillermo Kemper, Alonso Sanchez, Aldo Diaz, Jorge Gonzalez, Ruth Rubio-Noriega, Levy Boccato, Hugo E. Hernandez-Figueroa, *Senior Member, IEEE*

*Abstract*— Leaf-cutting ants, notorious for causing defoliation in commercial forest plantations, significantly contribute to biomass and productivity losses, impacting forest producers in Brazil. These ants construct complex underground nests, highlighting the need for advanced monitoring tools to extract subsurface information across large areas. Synthetic Aperture Radar (SAR) systems provide a powerful solution for this challenge. This study presents the results of electromagnetic simulations designed to detect leaf-cutting ant nests in industrial forests. The simulations modeled nests with 6 to 100 underground chambers, offering insights into their radar signatures. Following these simulations, a field study was conducted using a drone-borne SAR operating in the P-band. A helical flight pattern was employed to generate high-resolution ground tomography of a commercial eucalyptus forest.

A convolutional neural network (CNN) was implemented to detect ants and estimate their sizes from tomographic data, delivering remarkable results. The method achieved an ant nest detection accuracy of 100%, a false alarm rate of 0%, and an average error of 21% in size estimation. These outcomes highlight the transformative potential of integrating Synthetic Aperture Radar (SAR) systems with machine learning to enhance monitoring and management practices in commercial forestry.

*Index Terms*— Electromagnetic simulation, ant nest detection, SAR imaging, industrial forest pest monitoring.

## I. INTRODUCTION

CELLULOSE, extracted from the trunks of trees in industrial forests, is widely used in the paper, pharmaceutical, and textile industries. Brazil is one of the largest pulp producers in the world, with 9.6 million hectares of industrial forests [1,2]. *Pinus taeda* planted 2.5m apart in a square grid in just seven years can grow to 9m tall and 21.9cm in diameter. *Eucalyptus saligna* planted 3m apart can grow to 26.8m tall and 17.8 cm in diameter [20, 21]. Industrial forests face different pest threads depending on the region. Brazil's primary pest thread comes from leaf-cutting ants *Atta sexdens* and *Atta laevigata*. These ants can reduce the average Brazilian productivity by 14% annually [3], reaching 50% in some regions of the Brazilian Midwest without leaf-cutting ant control [4].

Leaf-cutting ants construct intricate subterranean nests equipped with chambers to provide optimal conditions for safeguarding and fostering their offspring. A single mature ant nest can contain over 7800 chambers, occupying an area of approximately 100 m² on the soil surface [22]. These chambers are spherical and measure between 12 to 25 centimeters in diameter, as seen in Fig. 1. Even at just two months old, these colonies can already cause damage to the forest and take up approximately 0.1 m² of surface area, with chambers up to 20 cm deep. A 3-year-old colony is considered mature, occupying a superficial area (called a loose soil mound) of about 100 to 200 m2, with chambers located below ground within the projection limit of the loose soil at depths between 0.5 to 7 meters [5, 22], as illustrated in Fig. 1.

Forest companies manage leaf-cutting ants by previously monitoring the surface area of their nests, which is measured in square meters [23]. Ant nest sampling covers 2-6 % of the cultivated area, and mathematical models estimate the total number of ant nests across all areas. Monitoring these nests involves visual detection by field workers [6]. However, hiring many workers to monitor large-cultivated areas and visually detect small nests is challenging, increasing the sampling error and affecting control decision-making accuracy [3]. Therefore, autonomous remote monitoring methods emerge as an alternative to these problems.

Synthetic Aperture Radar (SAR) systems are commonly deployed on satellites and airplanes. However, sizeable fixed-wing unmanned aerial vehicles (UAVs) have recently been used for SAR mapping [7]. The UAV-borne SAR systems offer superior spatial resolution and shorter revisit time than satellite or aircraft-based systems. This makes them ideal for low-altitude operations such as city surveillance, crop monitoring, and buried object detection. In addition, UAV-based systems can perform challenging flight paths like linear and circular tracks, surpassing those of satellite or aircraft-based systems. Circular and helical path techniques have been proven to offer excellent resolution for horizontal and vertical polarizations [8-10]. Furthermore, the penetration capability of radar signals has been extensively investigated across various scenarios. The results indicate this feature is influenced by soil properties and signal wavelength [11,12].

Tomographic SAR (TomoSAR) techniques can now generate 3D models from multiple 2D signals. This enables the segregation of scattering signals from different heights and their projection in the same range/azimuth resolution cell. The third dimension can be estimated more accurately by taking multiple parallel acquisitions of a single area, eliminating shadowing or layover phenomena [13,14]. In 2021, Minh *et al.* proposed a method for classifying different types of forests based on P-band tomographic SAR data obtained by using the F-SAR sensor of the German Aerospace Center [15]. They discovered that SAR tomography's high sensitivity to vertical forest structures and the Random Forest technique provided optimal results for forest classification. Another study by Ramachandran *et al.* examined the impact of polarization on forest height estimation using tomographic SAR processed with P-band data. The study found that the VV polarization in the P-band produced more precise results to estimate forest height [16]. In 2016, Schreiber utilized an L-band SAR system with a multi-static configuration to create three-dimensional images to detect landmines and unexploded ordnance (UXO) at depths up to 30 cm. While some false alarms were presented, the system could accurately identify targets on the surface and at this depth [17].

This paper introduces a novel technique to detect and measure the size of leaf-cutting ant nests under tree canopies. Tomographic SAR data collected by a drone-borne SAR system [18] were analyzed using Convolutional Neural Network (CNN) methods [19]. The SAR system followed a helical trajectory, which improved the resolution in the vertical direction. Full-wave electromagnetic simulations were also used to assess the feasibility of using Synthetic Aperture Radar (SAR). Utilizing SAR data to detect ant nests eliminates the need for manual sampling, significantly reducing labor costs. Furthermore, it ensures complete coverage of the mapped area (census).

The paper is structured as follows: Section 2 describes the characteristics of the industrial forest and the leaf-cutting ant nest, forming the foundation for the electromagnetic simulation. Section 3 details the overall setup of the electromagnetic simulation and the scenarios tested. Section 4 reports the results of flight surveys conducted at the study site. Section 5 provides a detailed explanation of the dataset and neural network. Section 6 focuses on the ant nest detection methodology, while Section 7 addresses the estimation of nest size. Section 8 discusses the findings, and Section 9 concludes the paper.

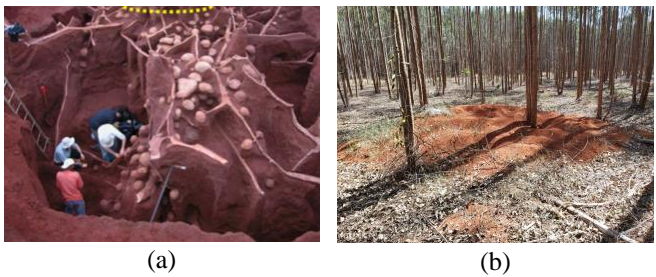

**Fig. 1.** (a) Internal structure showing the identifiable tunnels and chambers and (b) superficial view of a leaf-cutting ant nest.

## II. Methodology and results

### A. Ant Nest Numerical Modeling

A numerical study using numerical electromagnetic simulations deepened the understanding of how radar waves interact with ant nests. The P-band frequency range was chosen for its superior penetration capability compared to other brands in the drone's SAR system. Following the simulation, real-world data was analyzed to investigate this interaction further.

Full-wave numerical time-domain electromagnetic simulations were conducted using CST Design Studio 2022 [24] to evaluate the capability of SAR systems in detecting ant nests within a forested area surrounded by trees. The simulations used a plane wave and a P-band signal with radar-like characteristics – a central frequency of 425 MHz and a bandwidth of 50 MHz, as an excitation. The numerical setup included trees, sandy soil, and an underground ant nest. The trees were modeled as 10-meter-tall cylinders with a 25 cm diameter, a dielectric constant of 8, and a material density of 900 kg/m³ [25,26]. They were arranged 2.5 meters apart in all directions to create a symmetrical scenario.

As illustrated in Fig. 2, a grid of nine equally spaced trees was used to balance computational efficiency and accuracy. This allowed for analyzing the impact of ant nests on the central tree and its surrounding neighbors without significantly increasing computational costs. Material losses were neglected to further reduce computational demands. The soil layer was modeled with a precise thickness of 4.5 meters using the CST multi-layer tools.

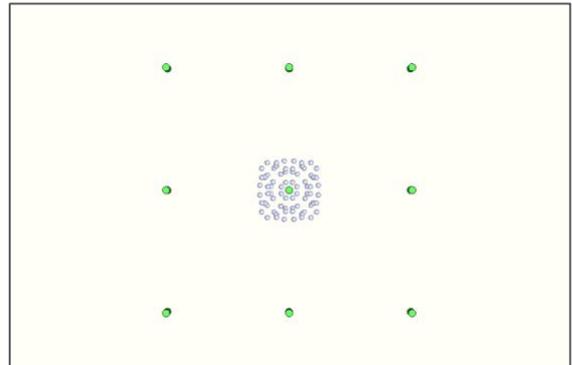

**Fig. 2.** Top-down view of the numerical model utilized to analyze the influence of ant nests and trees on the SAR signal. The model's green circles represent the trees, while the sphere symbolizes the ant nest chamber.

Since ant nests primarily comprise tunnels and chambers, the latter forming the bulk of their structure, the simulation focused solely on chambers. For simplicity, the nests were designed with octant symmetry. As detailed in Table I, the chambers were modeled as air-filled spheres with a dielectric constant of 1 and diameters ranging from 16 to 22 cm.

To make detection more challenging, the study used small ant nests with varying numbers of chambers, ranging from 6 to 100. These nests corresponded to surface areas between 0.6 m² and 10.2 m², as described by Pretto [27]. The CST probe tool was used to monitor the reflected waves within the area of interest, capturing critical data on radar interactions with the simulated nests.

This approach provided valuable insights into the electromagnetic behavior of radar signals interacting with leaf-cutting ant nests, setting the stage for improved detection and monitoring techniques.

SAR mapping leverages a helical flight path to achieve high-resolution imaging in the vertical direction, a critical feature for tomography. This technique eliminates vertical ambiguities and offers superior vertical and planimetric resolution compared to linear trajectories. Consequently, the tomographic data obtained from this method is of exceptional quality [8].

TABLE I
THE MAIN CHARACTERISTICS OF SIMULATED NESTS

| chambers number | 100 | 50 | 20 | 6 |
|---|---|---|---|---|
| chambers diameter (cm) | 22 | 20 | 18 | 16 |
| depth (m) | 0.4–1.5 | 0.3–1.0 | 0.3–0.6 | 0.2–0.4 |
| area (m²) | 10.24 | 6.25 | 2.25 | 0.64 |

The simulated helical flight trajectory was modeled using four circular flight paths with diameters ranging from 46 to 52 meters and altitudes between 27 and 40 meters. Numerical probes were strategically placed to align with these dimensions, ensuring an accurate replication of the helical flight and effective capture of reflected radar signals. These parameters were meticulously designed to replicate incidence angles comparable to those of the drone-based SAR system, enabling a comprehensive and realistic mapping of the simulated scenario.

The ant nests were centrally positioned within the simulation area, varying in the number of chambers (100, 50, 20, and 6), as illustrated in Fig. 3. Horizontal slices at different depths were taken to extract tomographic data, generating a series of SAR images for each mapped scenario. This approach ensured that the simulated data closely mirrored real-world conditions, enhancing the reliability and applicability of the results.

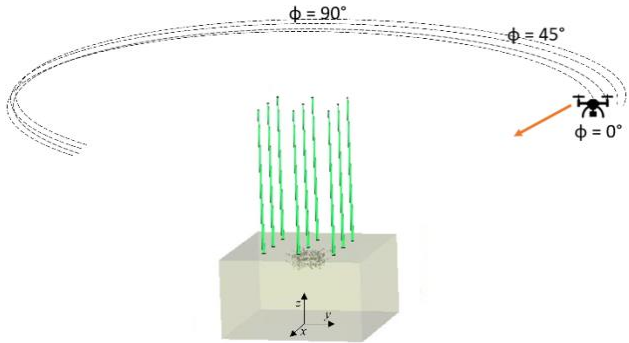

**Fig. 3.** Illustration of a helical mapping configuration featuring a nine-tree arrangement.

For SAR mapping based on helical flights, the electromagnetic simulation setup was rotated around the z-axis to emulate the movement of the SAR platform. At the same time, the probe tool and plane wave remained stationary. Leveraging the setup's octant symmetry, simulations were performed by rotating the configuration from 0° to 45°, and the data was mirrored to achieve a complete 360° mapping. The resulting data was processed using the back-projection algorithm outlined in [28] to generate SAR images.

Two scenarios were simulated: one featuring only trees and another incorporating trees and ant nests. The nests contained between 6 and 100 chambers, with an average depth of 0.5 meters. The primary goal was to evaluate the impact of ant nests on tree reflectivity. Full-wave electromagnetic simulations were processed to produce surface SAR images, from which reflectivity profiles were derived. These profiles were extracted along an east-to-west axis, intersecting the central tree in the simulated environment for detailed analysis.

Fig. 4 presents the resulting reflectivity profiles, showcasing three distinct peaks corresponding to the trees in this profile. This analysis highlights the impact of ant nests on the reflectivity signatures captured in the SAR images.

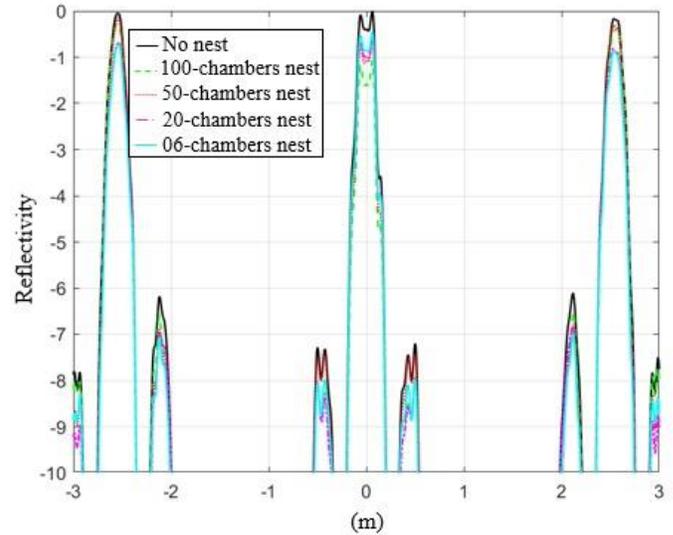

**Fig. 4.** Reflectivity profiles over axis $y$ for different scenarios, comparing cases with and without ant nests, mapped using a helical trajectory.

The reflectivity attenuation of the central peak, corresponding to the tree situated directly above the ant nests, increases proportionally with the nest size. While the side trees are also affected, their attenuation is less pronounced. For the nests beneath the central tree, reflectivity decreases by -0,34 dB for a 6-chamber nest and up to -1.98 dB for a 100-chamber nest.

Tomographic processing was applied to analyze subsurface features, generating surface ($xy$) SAR image layers from the surface to 0.5 meters underground. Reflectivity profiles derived from these layers were examined to assess the impact of the nests on the central tree's surface reflectivity. As shown in Fig. 4, the presence of ant nests causes a measurable attenuation in reflectivity compared to scenarios without nests.

Fig. 05 illustrates the reflectivity profiles for ant nests containing 6 and 20 chambers, analyzed across multiple horizontal layers ranging from z = 0 m to z = -2 m. The findings demonstrate the pronounced impact of ant nests on radar return signals. The 6-chamber nest exhibited the lowest reflectivity attenuation at a z = -0.4 m depth, while the 20-chamber nest reached its minimum attenuation at z = -0.6 m. Due to the diminishing focus on deeper horizontal layers in SAR imagery, the analysis was confined to a maximum depth of z = -2.

Full-wave electromagnetic simulations, conducted using parameters reflective of a real drone-borne SAR system, demonstrate that helical trajectory mapping is a highly effective method for locating ant nests. This approach outperforms alternatives due to its extended integration time and superior vertical resolution capabilities. When combined with tomographic processing, nest identification accuracy is further enhanced.

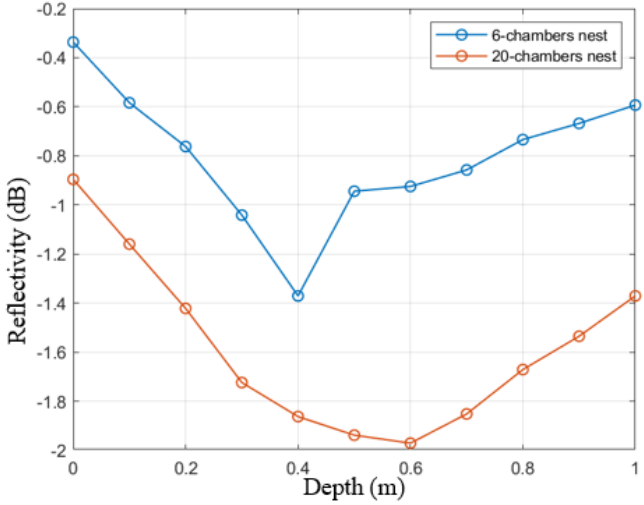

**Fig. 5.** Reflectivity for ant nest with 6 and 20 chambers from 0 to 2 m deep.

These numerical results validate the application of this technique in real-world scenarios. They offer a reliable and efficient solution for mapping and detecting ant nests in complex environments.

### B. Field SAR Surveys

The SAR surveys were conducted in a commercial plantation in the Ortigueira municipality of Parana State, Brazil (24°13'55"S and 50°47'34''W; at 750 m of altitude). The stand covers an area of 32.56 hectares and contains *Eucalyptus saligna* 5-year-old trees spaced 3.40 x 1.75 m apart. Such an area transitions between the Atlantic Forest and Cerrado biomes. The Köppen classification identifies the climate as subtropical without a dry season and mild summers. The mean yearly temperature is 18.6℃, and the average rainfall is 1443 mm [29,30]. Leaf-cutting ants, including *Atta sexdens* and various species of *Acromyrmex spp*, are present within this region.

A ground survey was carried out to identify all the leaf-cutting ant nests. Eighty nests of varying sizes were located, and their positions were recorded using a differential GNSS for reference data. The yellow pins in Fig. 6 mark the locations of these nests.

The multiband drone-borne SAR system RD350 Explorer was used to map the area. The system illustrated in Fig. 7 simultaneously operates in three different frequency bands, L, P, and C, enabling it to acquire complete multi-band information with a single survey [18]. Table II presents the RD350 Explorer's main specifications. A helical path was selected due to its superior resolution in horizontal and vertical dimensions and its efficacy in suppressing sidelobes. Figure 8a illustrates the conical orbit, which has variable radii ranging from 115 to 165 meters and heights from 120 to 80 meters. This flight path effectively illuminates a subsoil volume measuring 100 by 100 square meters and covering a depth of 10 meters

[8]. The drone-borne SAR conducted five helical flights to map five distinct regions of the study site, as illustrated in Fig. 8b.

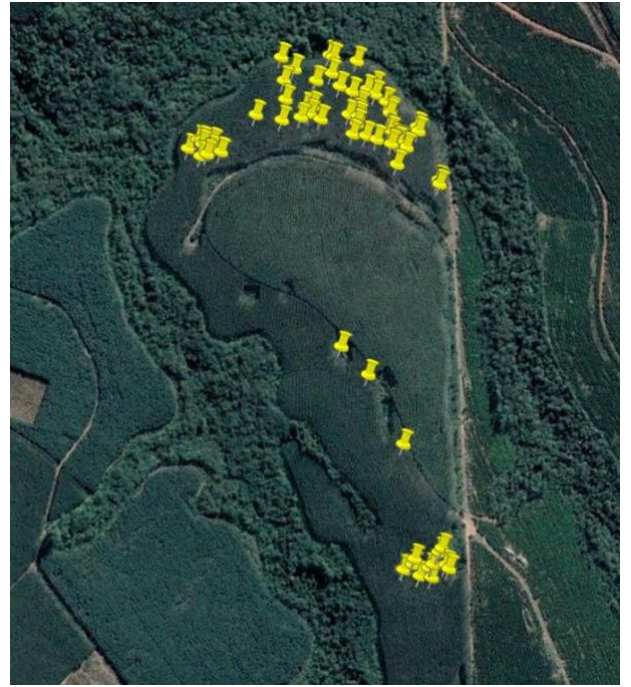

**Fig. 6.** Study site with yellow markers highlighting the locations of ant nests.

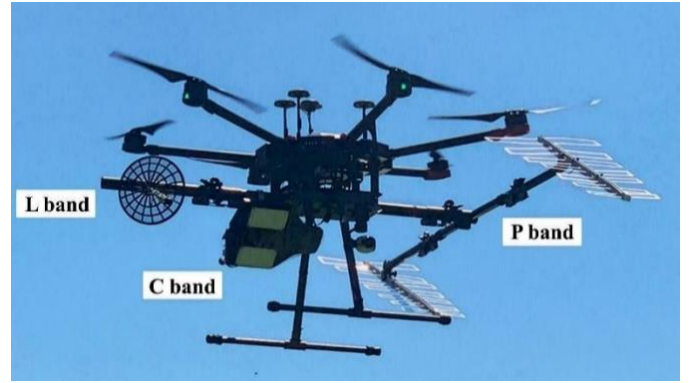

**Fig. 7.** Multiband drone-borne SAR system RD350 Explorer.

TABLE II
RD350 EXPLORER PARAMETERS

| Radar Parameters | P band | L band | C band |
|---|---|---|---|
| Polarization | HH | HH | VV |
| Wavelength (cm) | 70.5 | 22.8 | 5.6 |
| Bandwidth (MHz) | 50 | 150 | 200 |
| Azimuth aperture (°) | 55.9 | 58.5 | 32.5 |
| Elevation aperture (°) | 69.3 | 79.8 | 51.3 |
| Azimuth resolution (cm) | 30.0 | 10.0 | 5.0 |
| Range resolution (m) | 4.0 | 1.2 | 1.0 |

The P-band data was processed using a time-domain back-projection algorithm. For each region, eight synthetic aperture radar (SAR) images were generated, accurately depicting an area of 100 x 100 m² from the surface down to a depth of 2.1 m. The SAR images underwent radiometric correction to ensure accuracy, considering the distance between the platform and the illuminated target, the bidirectional antenna pattern, and the number of pulses used to construct the synthetic aperture through the back-projection algorithm. When flying along circular or helical paths, the planimetric resolution can be estimated to be about one-quarter of the wavelength. For the P-band Synthetic Aperture Radar (SAR), the expected resolution is 0.18 meters.

Figure 9 presents P-band images of Region 4, one of the five areas selected for validation. The high surface resolution of the helical flight allows one to identify the rows of trees in the industrial plantation.

Electromagnetic simulation results reveal that ant nests reduce the reflectivity of the surrounding volume. Leveraging this characteristic, a specific reflectivity threshold was applied to generate the 3D images shown in Fig. 10. This figure provides a SAR tomography of a fieldwork site where two ant nests were identified.

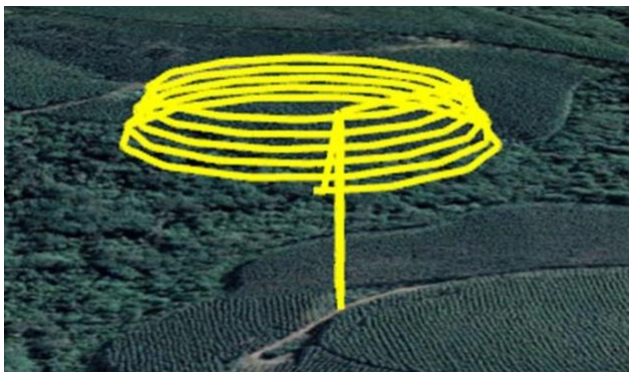

(a)

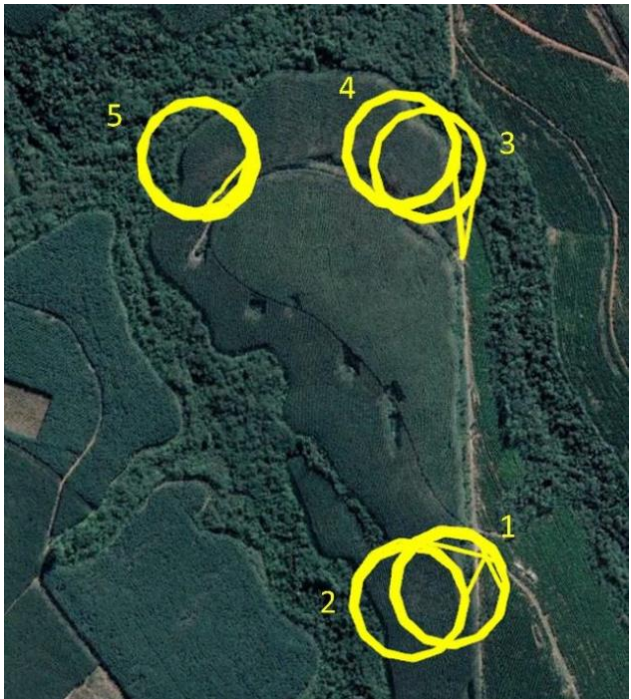

(b)

**Fig. 8.** (a) Perspective view of a helical flight path. (b) Top view of five helical flights over five distinct regions within the study site.

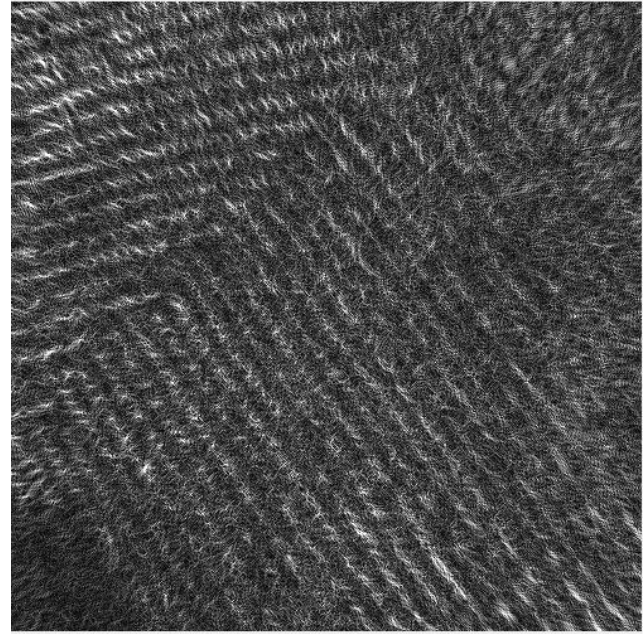

**Fig. 9.** SAR image of Region 4 showing details of the rows of trees in the mapped area.

For the first time, Fig. 10 presents a tomographic representation of two nests belonging to the *Atta genus*, depicted in both top and perspective views. The smaller nest (#2) extends to a depth of approximately 1.3 m, while the larger nest (#1) reaches about 2 m deep. Previous studies indicate that the maximum depth of nest chambers correlates with nest surface area [22], with manually excavated *A. sexdens* nests reaching depths of 2 to 3 m. However, most chambers are concentrated between 0.5 and 2 m [27], consistent with our findings. In the 3D volumes, the low-reflectivity regions correspond to the ant nests, while the high-reflectivity voids represent the surrounding soil structure. This visualization offers valuable insight into their natural habitat's spatial organization and depth.

The 3D subsurface images presented in Fig. 10 were generated using the following methodology:

**Layer Generation:** Eight surface layers were constructed using the back-projection algorithm. These layers start 30 cm below the surface, with 30 cm intervals between each layer, reaching a maximum depth of 2.10 m.

**Thresholding:** The voxels were assigned a reflectivity threshold. Voxels with reflectivity values below the average reflectivity of standard soil were assigned a value of 1, while all others were assigned a value of 0.

**Interpolation:** A three-dimensional interpolation between the voxels created two blue volumes. These volumes represent regions with reflectivity values lower than the average reflectivity of the surrounding standard soil.

This approach effectively highlights areas of reduced reflectivity, corresponding to the subsurface ant nest structures. The total estimated surface area of ant nest #1 is approximately 105 m², compared to an estimated 115 m² measured during fieldwork. The slightly larger area observed in the field aligns with expectations, as most nest chambers typically lie within the surface projection boundaries of the nest [22].

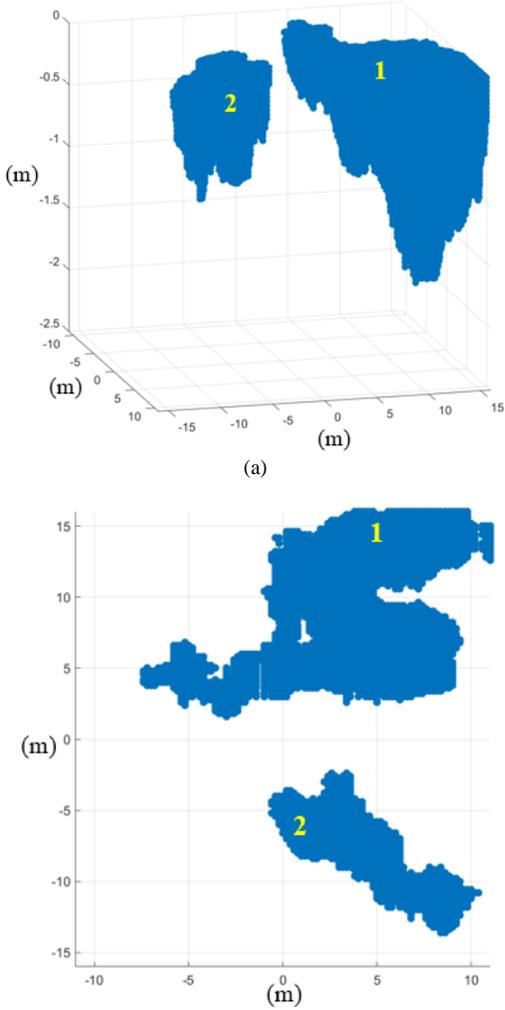

**Fig. 10.** Tomographic image of two ant nests. (a) Perspective view. (b) Top view.

## C. Ant Nest SAR Mapping

The ant nest mapping method proposed in this study consists of two distinct stages: detection and surface area estimation of the nests. Both stages leverage standard SAR imagery, specific reference data, and convolutional neural networks (CNNs).

A 100 x 100 m² reference image was generated for each mapped region. These images depict the ant nests identified during fieldwork, represented as circular areas centered on their locations and approximate sizes.

The precise shapes of the nests remain unknown. Fig. 11 illustrates the reference images for the five regions, highlighting the nests mapped in the study.

When used for detection, the reference images were designed with a pixel value of 0 for the background and 1 for the circular areas. Alternatively, the circular areas were assigned pixel values corresponding to the nest sizes for size estimation. Fig. 12 illustrates the profiles of reference images configured for detection and size estimation.

To prepare the data, both SAR and reference images were divided into sub-images, each measuring 12 x 12 m² with a 4 m stride. This process yielded 2,420 SAR sub-images with dimensions of 61 x 61 x 8 pixels and corresponding reference sub-images with dimensions of 61 x 61 pixels. The average pixel value within the central region of each reference sub-image was computed to determine the probability of detection or nest size. These calculated values served as reference data for their corresponding SAR sub-images.

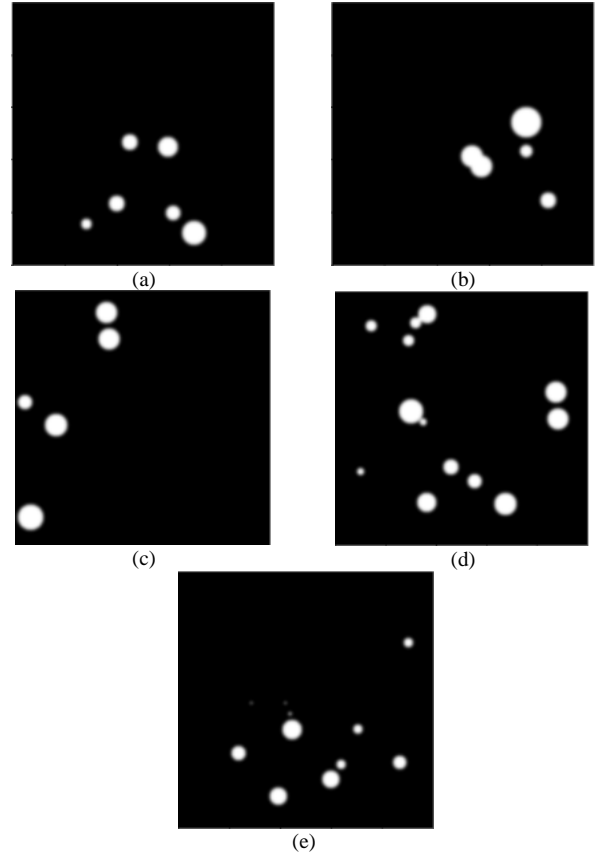

**Fig. 11.** Reference images for the five regions are shown, with white circles indicating the ant nests identified during fieldwork: (a) Region 1. (b) Region 2. (c) Region 3. (d) Region 4. (e) Region 5.

For model development, SAR sub-images from Regions 1 to 4 were used for training, while those from Region 5 were reserved for validation.

The dataset comprised 1,936 SAR images and 484 SAR sub-images, each paired with corresponding reference data, for training and validation.

The study included 40 ant nests, 29 allocated for training and 11 for validation. Table III provides additional details about the training and validation datasets.

Fig. 13 shows the schematic of the Convolutional Neural Network (CNN) used for monitoring ant nests with SAR sub-images. The same architecture, with minor adjustments discussed later, was employed to detect ant nests and estimate their size. The CNN was designed as a regression model for size estimation, with a single neuron in the output layer.

The input to the CNN consists of SAR sub-images with dimensions of 61 x 61 x 8 (height, width, and channels). The corresponding reference data includes detection probabilities or size information for each sub-image.

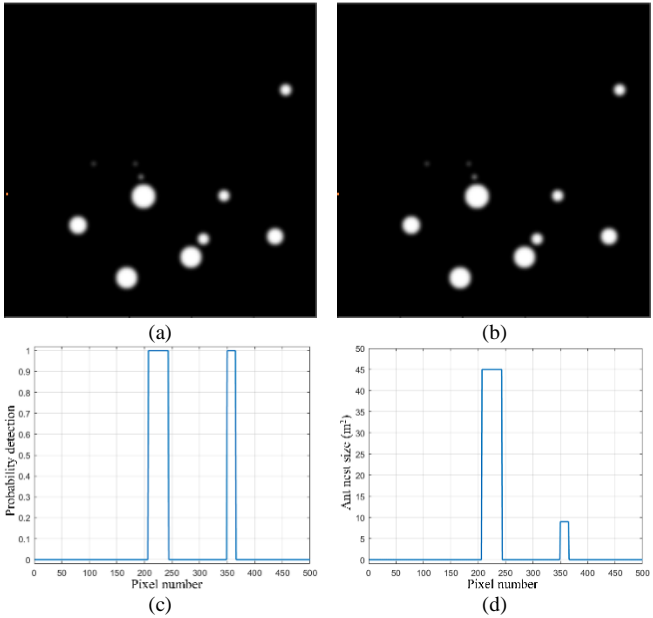

**Fig. 12.** Reference image for (a) detection and (b) size estimation in which orange dotted lines are drawn to obtain profile information for (c) probability detection and (d) size estimation.

The architecture features two consecutive convolutional layers with 64 filters, 3 x 3 kernels, Batch Normalization, and Max Pooling. This structure is repeated with 128 filters and two additional convolutional layers with 256 filters and 3 x 3 kernels. The final layers include a fully connected layer and an output layer representing detection probability or size estimation. All hidden layers use the ReLU activation function.

The network was trained from scratch without leveraging any pre-trained models. It was initialized with random weights and parameters, and the design was fine-tuned empirically. Based on performance tests with various CNN configurations, key hyperparameters, such as kernel size, the number of filters, and the structure of convolutional and dense layers, were optimized.

TABLE III
DETAILED INFORMATION OF THE DATASETS.

| Dataset | Number of ant nets considered | Number of sub images | Number of sub images containing ant nest |
|---|---|---|---|
| Training | 29 | 1936 | 117 |
| Validation | 11 | 484 | 27 |

Fig. 14 provides an overview of the ant nest monitoring process. The input consists of 8 SAR images, representing a 100 x 100 m² area captured at different depth slices. Georeferenced SAR sub-images are then fed into the respective CNN models. In the example illustrated, the detection CNN outputs a value of 1, signifying the highest probability of nest presence, while the size estimation CNN outputs a value of 20, indicating an estimated nest size of 20 m². These numerical outputs are subsequently used to generate output images. For detection, the output image marks the exact location corresponding to the SAR sub-image input. For size estimation, it highlights a circular area equivalent to 20 m². This process is repeated for every SAR sub-image, systematically creating a comprehensive map of ant nest locations and sizes.

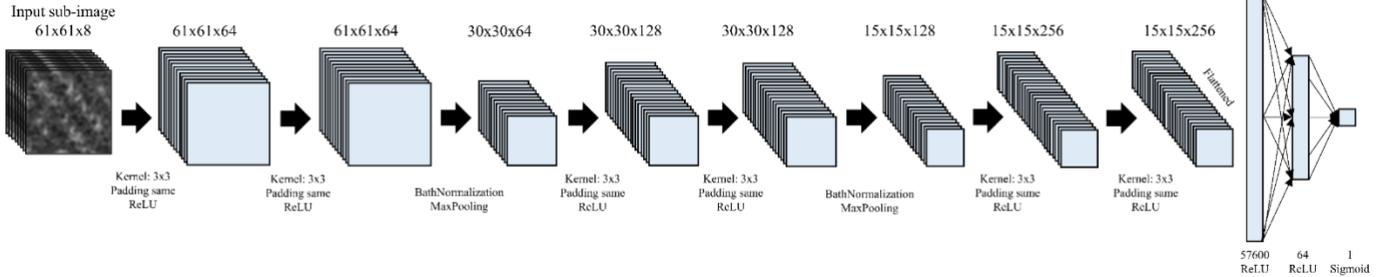

**Fig. 13.** Convolutional Neural Network (CNN) architecture designed for ant nest detection using SAR imagery.

## D. Ant Nest Detection

A Convolutional Neural Network (CNN) was developed to detect ant nests using the study site's Synthetic Aperture Radar (SAR) images. The CNN parameters were optimized through training and validation datasets. Data augmentation techniques enhanced detection accuracy, including horizontal and vertical flips, brightness adjustments, and rotations.

The CNN's output layer employs the Sigmoid activation function, producing a probability score for detecting an ant nest. The binary cross-entropy loss function was utilized alongside the Adam optimizer, a stochastic gradient-based optimization method for training. The learning rate was set to 0.00003, and the model was trained over 35 epochs with a batch size of 64. Fig. 15 displays the accuracy curves for both training and validation datasets.

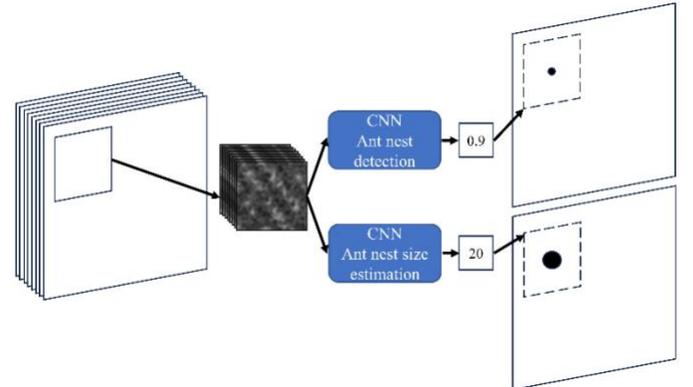

**Fig 14.** Overview of the process for ant nest monitoring.

The training accuracy curve demonstrates a smooth learning progression, while the validation curve reflects variations typical of smaller datasets. Together, these curves highlight the model's stability and generalization performance during the training process.

The performance of the adjusted CNN model with the highest accuracy was evaluated for detecting ant nests in SAR images across five mapped regions. Georeferenced SAR sub-images, aligned by their central pixel, were used as inputs to the CNN, which output the probability of ant nest detection for each sub-image. Using these probabilities, each region generated an ant nest position map with the same dimensions as the SAR image, as outlined in Fig. 13.

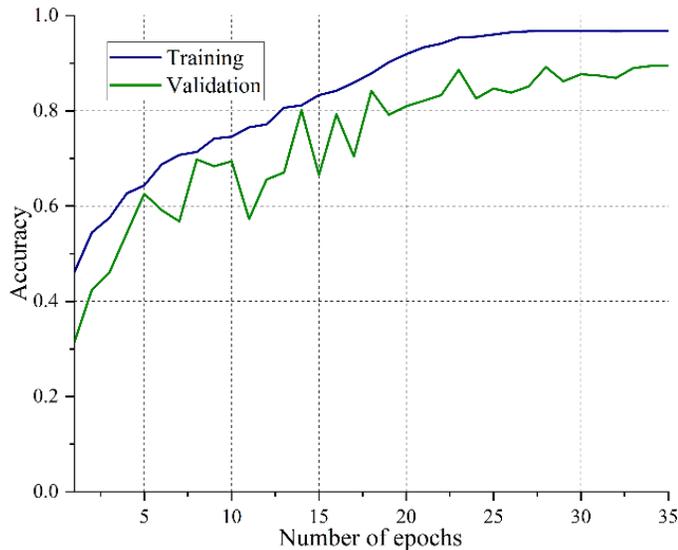

**Fig. 15.** Accuracy plot for the training and validation datasets.

Fig. 16 compares estimated and actual ant nest positions for each region. The yellow circles indicate the ground-truth locations of 29 ant nests identified during fieldwork, while the purple dots represent the CNN-estimated positions. These results were derived from the training dataset. On the map, each yellow dot, denoting an actual ant nest, is encircled by clusters of purple points, signifying perfect alignment without any planimetric error (discrepancy between actual and estimated positions). Notably, no purple points appear outside the yellow circles, indicating a 100% detection rate with zero false alarms. This demonstrates the model's exceptional accuracy and reliability in identifying ant nest locations.

Fig. 17 displays the locations of ant nests in Region 5 as estimated by the CNN. This region was selected for validation due to its diversity of nest sizes, ranging from 1 to 50 m², which adds complexity to the prediction process. Validation results revealed a planimetric error, detailed in Table IV, with a standard deviation of 5.6 m. Several factors may contribute to this discrepancy:

(a) **Ground Truth Identification Method:** The ground-truth nest locations were determined through visual inspection, focusing on the center of loose soil mounds. However, this method does not provide precise information about the nests' subsurface structures. For instance, some underground chambers of *Atta sexdens* nests may extend beyond the surface projection limits of the loose soil, creating a misalignment between the chamber center and the mound.

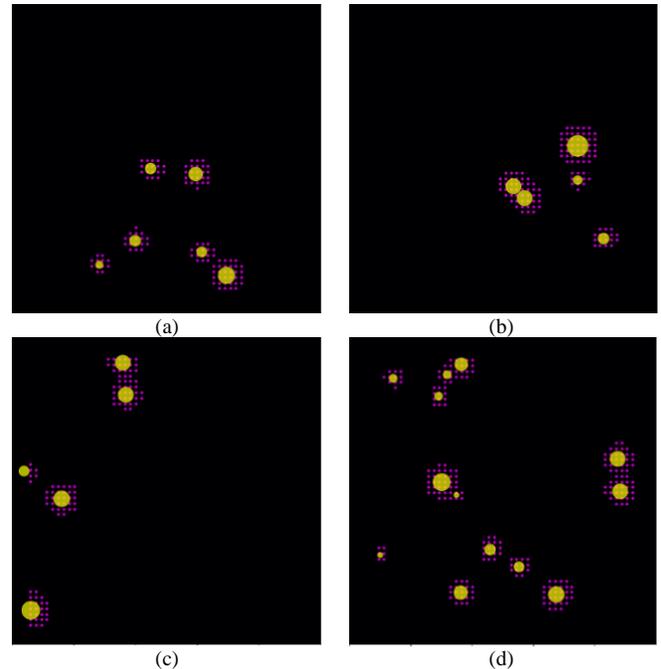

**Fig. 16.** Comparison of estimated and actual ant nest positions: (a) Region 1, (b) Region 2, (c) Region 3, and (d) Region 4.

(b) **Irregular Surface Features:** *Atta sexdens* nests often feature multiple mounds of loose soil scattered irregularly on the surface, complicating the identification of the nest's central location.

(c) **GNSS Accuracy Limitations:** The differential GNSS station used in this dense forest environment offers a position accuracy of up to 0.5 m, which could also contribute to minor positional discrepancies.

Despite these challenges, the CNN demonstrated robust performance in detecting nest locations, with errors primarily attributed to the limitations of ground-truth data collection and the inherent complexity of the nests' surface features.

Based on the radar-detected locations and their corresponding ground-truth data, the model achieved a 100% detection rate with 0% false alarms and a location standard deviation error of 5.6 m. Notably, fieldwork revealed no loose soil in the northwest area of Region 5, and consistent with this observation, the radar also did not detect any ant nests in that area. This alignment underscores the accuracy and reliability of the detection methodology.

The radar detection radius was determined based on the maximum detection distance observed in Fig. 16, with an additional 30% margin added to account for the absence of ground truth data. Specifically, the maximum distance in Fig. 16—19 meters between radar detection position 8* and the nearest loose soil (position 2)—was increased by 30%, resulting in a final detection radius of 25 meters. In practical terms, this ensures that the ground operator has clear visibility of loose soil within a 25-meter radius from their position.

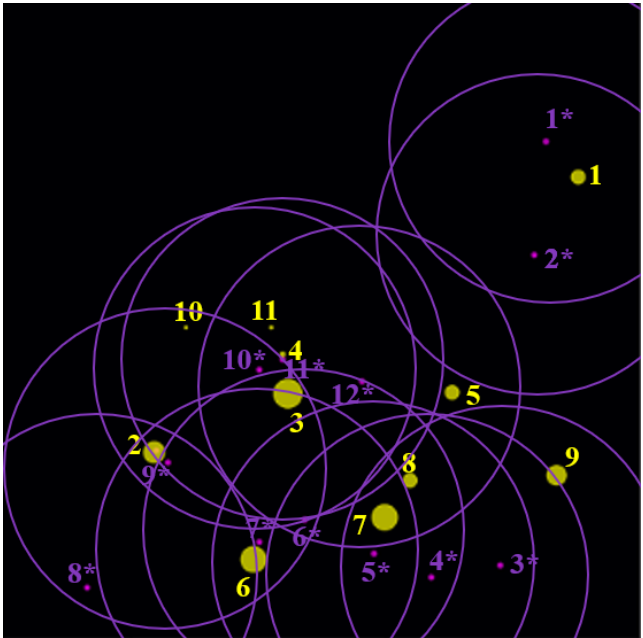

**Fig. 17.** Radar-detected positions (purple dots) and ground-truth locations (yellow dots) of ant nests in Region 5.

TABLE IV
RADAR, RELATED GROUND TRUTH, AND ERROR LOCATIONS

| Radar Detection | Related Ground Truth | Planimetric error (m) |
|---|---|---|
| 1* | 1 | 6.0 |
| 2* | 1 | 11.0 |
| 3* | 9 | 13.3 |
| 4* | 7, 8 | 9.5 |
| 5* | 7, 8 | 4.0 |
| 6* | 6, 7 | 8.5 |
| 7* | 6 | 1.5 |
| 8* | 2 | 19.0 |
| 9* | 2 | 1.5 |
| 10* | 3, 4, 10, 11 | 3.5 |
| 11* | 3, 4, 11 | 0.5 |
| 12* | 3, 4, 5 | 9.5 |

## E. Ant Nest Size Estimation

The ant nest size estimation method leverages a Convolutional Neural Network (CNN) and incorporates both SAR imagery and the detection results from the earlier stage. A CNN architecture like the one described in Subsection D and illustrated in Fig. 12 was employed, with a critical difference: the output layer uses a ReLU activation function to produce a positive real number corresponding to the estimated nest size.

The CNN weights were optimized using training and validation datasets to ensure accurate size predictions. The training used the mean squared error (MSE) loss function and the Adam optimizer, with a learning rate of 0.00003. The network was trained over 40 epochs with a batch size of 64.

The combined results from the CNNs for detection and size estimation provided accurate size estimates for the ant nests. The complete methodology, integrating detection and size estimation, is summarized in the block diagram presented in Fig. 18. The CNN algorithm generates an image that visually represents objects' estimated size, with white areas indicating intensity and size based on the CNN's output. However, as illustrated in Fig. 18, the size estimation results occasionally include false alarms.

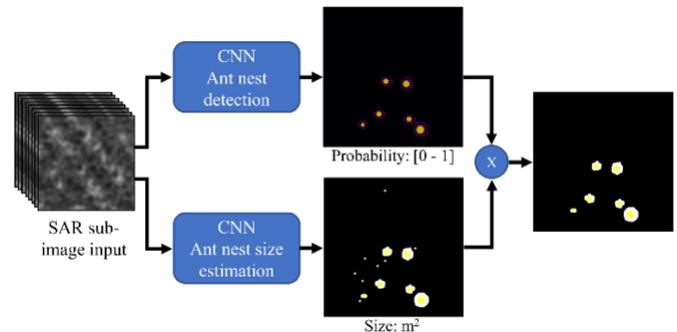

**Fig. 18.** Block diagram for ant nest size estimation considering the two CNNs.

The estimated results are refined by leveraging the ant detection results to address this issue. This is achieved by pixel-wise multiplication of the detection and size estimation images, effectively eliminating false positives. This approach was applied to both the training and validation datasets. After refinement, the estimated and measured sizes of ant nests are compared in Fig. 19, demonstrating the effectiveness of this method in improving accuracy.

The training dataset produced the most accurate estimates of ant nest sizes, ranging from 5 to 110 m². Interestingly, regardless of the dataset, the smallest nests consistently exhibited the highest estimation errors. In contrast, the validation dataset showed a mean error of 21.7%, approximately five times greater than the mean error in the training dataset. This discrepancy is primarily due to the more significant proportion of smaller nests in the validation dataset, making accurate size estimation more challenging.

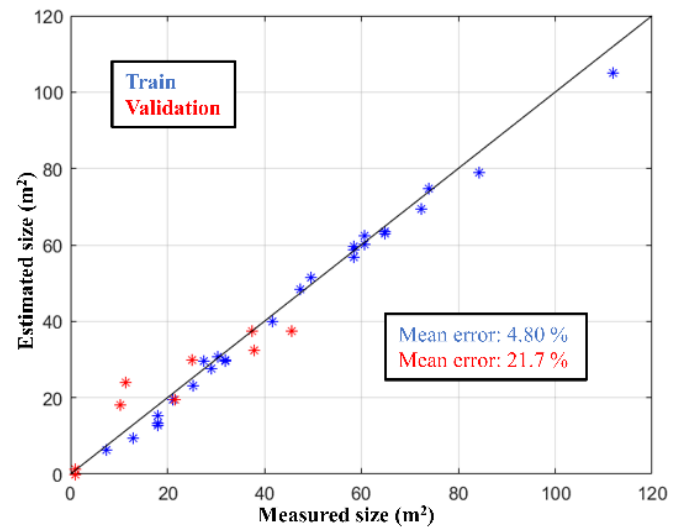

**Fig. 19.** Comparison of measured and estimated ant nest sizes for the training and validation datasets.

Performance metrics further highlight these differences. The training dataset achieved an $R^2$ of 0.98, an adjusted $R^2$ of 0.98,

and a root mean square error (RMSE) of 2.75 m², indicating exceptional accuracy. Meanwhile, the validation dataset achieved an $R^2$ of 0.76, an adjusted $R^2$ of 0.70, and an RMSE of 5.97 m². These results demonstrate the method's effectiveness in the training dataset while highlighting the challenges posed by smaller nests in the validation dataset.

While the fieldwork data on ant nest sizes is collected as surface area measurements (2D), it is utilized to train the CNN, which processes radar tomographic images containing 3D information. Ant nests are inherently 3D volumetric structures that appear in the radar tomographic data. The neural network analyzes these images to identify 2D patterns within the 3D data, enabling it to estimate the nest sizes accurately through its output.

## III. Discussion

An additional ant nest monitoring experiment was conducted to validate the robustness of the CNN-based methodology and confirm that the number of ant nests is sufficient for effective training. In this experiment, Region 2 was used for validation, while the other regions were used for training. The results are presented in Fig. 20.

Region 2, in contrast to Region 5, contains bigger ant nests, including the largest nest in the study area. Validation results showed a 100% detection rate with a 0% false alarm rate and an average planimetric error of 7 m. For size estimation, the training dataset yielded a mean error of 7.53%, while the validation dataset showed a mean error of 13.2%. These outcomes closely align with those from the initial experiment, demonstrating that the selected CNN model effectively addresses the problem.

The inclusion of 3D radar information proved critical for achieving accurate results. Additionally, a linear flight was conducted over the study area to assess the feasibility of using linear flight paths for ant nest monitoring. The detection results from 2D images captured during the linear flight mapping are presented in Fig. 21, highlighting the comparative performance of linear flight paths for this task.

Fig. 21a presents the SAR image captured during a linear flight, which lacks 3D information. The image is divided into training and validation, from which sub-images were extracted as previously described. Red circles indicate the locations of ant nests identified during fieldwork, following a presentation style like Fig. 16.

The results from the training and validation datasets highlight significant limitations, with numerous false alarms and very few accurately identified ant nests. This poor performance underscores the critical importance of the detailed 3D information provided by helical flights for effective and reliable ant nest monitoring.

An additional industrial forest area was surveyed using the drone-borne SAR system. Unlike the initial study area, this region featured smaller ant nests ranging in size from 0.05 to 10 m². Seven separate areas were mapped, producing seven sets of SAR images, each covering a 100 x 100 m² region and incorporating tomographic data.

Figure 22 presents the ant nest detection results for the second validation area near Lençóis Paulista, São Paulo State (22° 35' 46'' S, 48° 48' 40'' W; altitude 550 m).

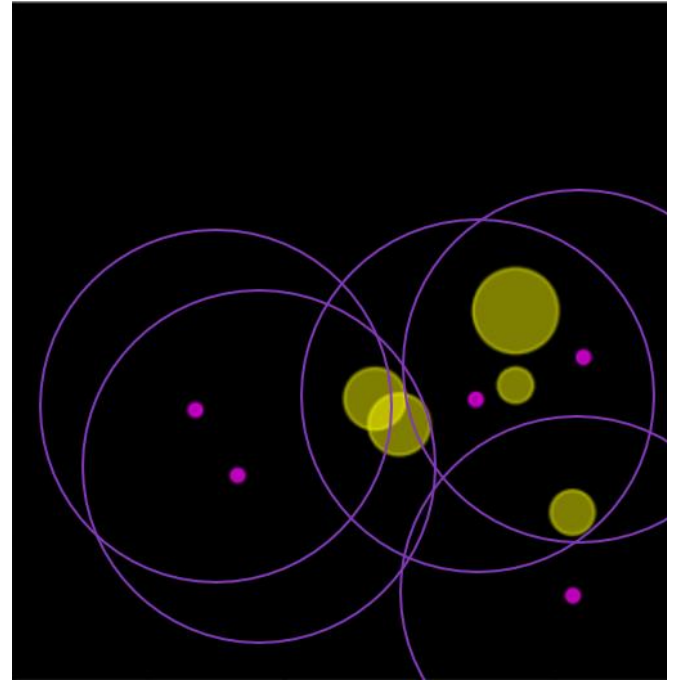

(a)

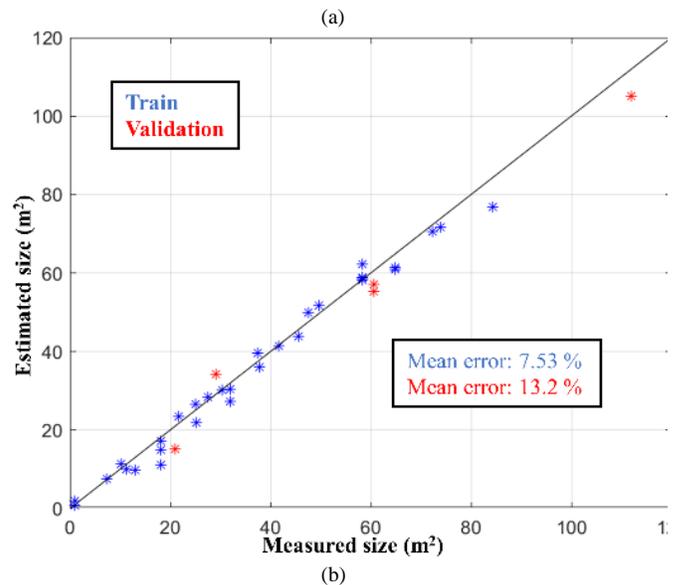

(b)

**Fig. 20.** Results of (a) ant nest detection in Region 2 and (b) size estimation in the training and validation datasets, considering Region 2 as validation and the rest as training.

The color scheme used for the detection results is consistent with that of Figure 14. The detection process faced the greatest challenge with the smallest central nests, achieving a 100% detection rate but with an average planimetric error of 10 meters. Notably, the false alarm rate remained at 0%. The maximum detection distance between the radar detection position and the nearest loose soil, as shown in Figure 21, was 25 meters, slightly exceeding the distance observed in Figure 16. These findings underscore the difficulty of detecting smaller nests in this specific area.

These findings represent a significant advancement in monitoring leaf-cutting ant nests. First, the method enables detection beneath dense tree canopies, a feat unattainable with conventional RGB imaging techniques. Second, it provides a level of accuracy that surpasses traditional manual nest size measurements [32].

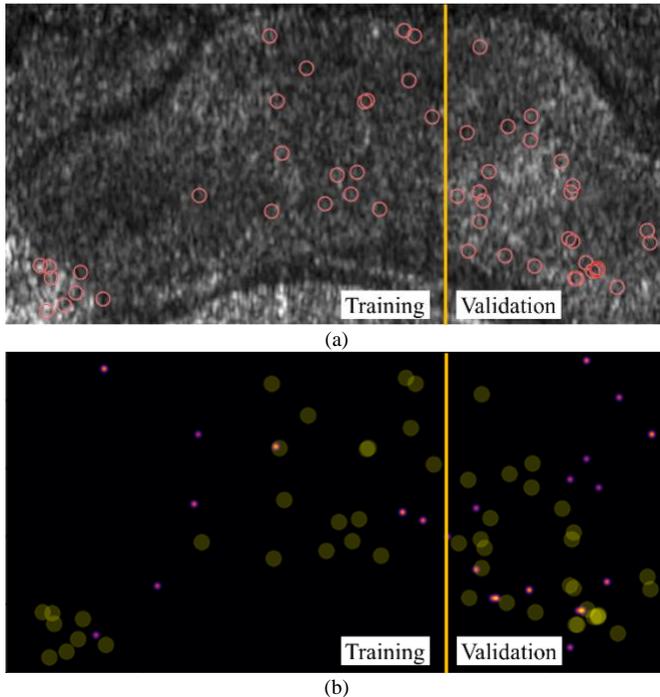

(a)

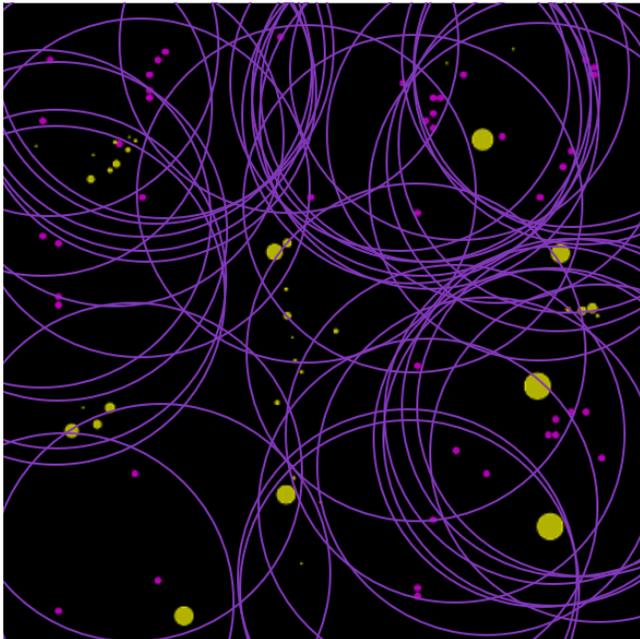

(b)

**Fig. 21.** (a) Linear flight SAR image segmented into training and validation areas. (b) Ant nest detection results are derived from the linear flight SAR image.

**Fig. 22.** Estimated ant nests (purple) positions compared with actual positions (yellow) in the second validation area.

Additionally, the radar system drastically reduces monitoring time compared to fieldwork. The helical flight path of the SAR system takes approximately 10 minutes, while tomographic processing using the back-projection algorithm and nest size estimation via neural networks was completed in roughly 30 minutes.

The back-projection processor used for imaging did not account for multi-angle backscatter variations; instead, it assumed a constant radar cross-section for targets, regardless of elevation or azimuth angles. Only the range-induced phase shift between the target and radar was considered [8]. However, this study incorporates electromagnetic simulations that include multi-angle backscatter information for ant nests. When processed with the back-projection algorithm, the raw data demonstrated effective detection capabilities, highlighting the robustness of the approach.

## VI. CONCLUSION

This study demonstrates the effective use of machine learning to detect and estimate the size of leaf-cutting ant nests using tomographic Synthetic Aperture Radar (SAR) images. The drone-borne SAR system provides a significant advantage, offering the flexibility to execute complex flight patterns, such as helical paths, which are critical for achieving the resolution required to monitor ant nests accurately.

The findings highlight the effectiveness of P-band SAR images acquired during helical flights in detecting ant nests within industrial forests. In the validation dataset for Ortigueira, the CNN achieved a 100% detection rate with 0% false alarms and a planimetric error standard deviation of 5.6 m. For size estimation, the mean error was 4.8% for the training dataset and 21.7% for the validation dataset. In the second validation area in Lençóis Paulista, the detection rate reached 87%, demonstrating the robustness of the approach across different conditions. Unlike traditional monitoring techniques, which sample only 2–6% of the total cultivated area, this method enables comprehensive mapping of the entire area of interest.

The adjusted CNNs efficiently processed SAR data within three hours, a dramatic improvement compared to the five days required for manual fieldwork and nest identification. This efficiency makes SAR-based monitoring a practical solution for large and remote areas where fieldwork is challenging. SAR mapping enables forest companies to detect leaf-cutting ant nests with unprecedented accuracy, outperforming conventional sampling-based techniques. By incorporating variability in nest sizes, this technology also facilitates more targeted and cost-effective application of insecticide baits in forest plantations.

Future developments will focus on reducing planimetric error to less than 5 meters for even greater positional accuracy. Efforts will also aim to refine nest size estimation, targeting a mean error of under 10%. Additionally, expanding the scope to include a wider variety of ant species will enhance the versatility and applicability of this approach, further solidifying its value in forest management.

## ACKNOWLEDGMENT

This work was partially supported by the São Paulo Research

Foundation (FAPESP) under the projects 2021/06506–0 (StReAM), 2021/11380–5 (CPTEn), and 2021/00199–8 (SMARTNESS); the Brazilian Agency CNPq, under the project 312714/2019–0 (HEHF's research productivity grant).

## REFERENCES

[1] J"Brazilian tree industry: statistical data" https://iba. org/eng/statistical-data (webpage accessed on Jan. 4th, 2023).

[2] J. Buongiorno, et al., "Outlook to 2060 for Worldforest and Forest Industries," General Technical Report SRS–151. Asheville NC, US.

[3] Zanetti, R., Zanuncio, J.C., Santos, J. C., Da Silva, W. L. P., Ribeiro, G. T., Lemes PG (2014) An overview of integrated management of leaf-cutting ants (Hymenoptera: Formicidae) in Brazilian forest plantations. Forests 5:439–454. doi 10.3390/f5030439]

[4] "Embrapa: pragas em plantios florestais produtivos," https://www.embrapa.br/florestas/transferencia-de-tecnologia/ pragas-florestais/perguntas-e-respostas/ (webpage in Portuguese, accessed on Jan. 4th, 2023).

[5] A. C. Swanson et al., "Welcome to the Atta world: a framework for understanding the effects of leaf-cutter ants on ecosystem functions," Functional ecology, vol. 33, no. 8, p. 1386 – 1399, 2019.

[6] R. Zanetti, W. SILVA, "Monitoramento de insetos-praga em plantações florestais," In: Novo Manual de Pragas Florestais Brasileiras.1 ed. Montes Claros: Instituto de Ciências Agrárias da Universidade Federal de Minas Gerais, 2021, pp. 53–91.

[7] C. Li, H. Ling, "Synthetic aperture radar imaging using a small consumer drone," IEEE International Symposium on Antennas and Propagation, Vancouver, BC, Canada, July 2015, pp. 685 – 686.

[8] J. Goes, "Techniques for High-Resolution 3D Images with Synthetic Aperture Radar," Ph.D. thesis, School of Electrical and Computer Engineering, University of Campinas, Campinas, SP, Brazil, 2022.

[9] O. Ponce, P. Iraola, M. Pinheiro, M. Rodriguez-Cassola, R. Scheiber, A. Reigber, A. Moreira, "Fully Polarimetric High-Resolution 3-D Imaging with Circular SAR at L-band," IEEE, Trans. Geosci. Remote Sens., vol. 52, no. 6, pp. 3074–3090, 2014.

[10]D. Feng, D. An, L. Chen, X. Huang, Z. Zhou, "Multicircular SAR 3-D Imaging Based on Iterative Adaptive Approach," ASPAR, Xiamen, China, November 2019, pp. 1–5.

[11]M. El Hajj, N. Baghdadi, H. Bazzi, M. Zribi, "Penetration Analysis of SAR Signals in the C and L Bands for Wheat, Maize, and Grasslands," Remote Sensing, vol. 11, no. 1, 2018, pp. 1–14.

[12]P. Paillou, Y. Lasne, E. Heggy, J. Malezieux, G. Ruffie, "A study of P-band Synthetic Aperture Radar Applicability and Performance for Mars Exploration: Imaging Subsurface Geology and Detecting Shallow Moisture," Journal of Geophysical Research, vol. 111, no. E6, 2006, pp. 1–12.

[13]S. Khoshnevis, S. Ghorshi, "A tutorial on tomographic synthetic aperture radar methods," SN Applied Sciences, 2020, pp. 1–14.

[14]C. Rambour, A. Budillon, A. Johnsy, L. Denis, F. Tupin, G. Schirinzi, "From Interferometric to Tomographic SAR: A Review of Synthetic Aperture Radar Tomography-Processing Techniques for Scatterer Unmixing in Urban Areas," IEEE Geoscience and Remote Sensing Magazine, vol. 8, no. 2, 2010, pp. 1–24.

[15]D. Minh, Y. Ngo, T. Le, "Potential of P-Band SAR Tomography in Forest Type Classification," Remote Sensing, vol. 13, no. 4, 2021, pp. 1–16.

[16]N. Ramachandran, S. Saatchi, S. Tebaldini, M. Alessandro, O. Dikshit, "Evaluation of P-Band SAR Tomography for Mapping Tropical Forest Vertical Backscatter and Tree Height," Remote Sensing, vol. 13, no. 8, 2021, pp. 1–22.

[17]E. Schreiber, M. Peichl, S. Dill, A. Heinzel, F. Bischeltsrieder, "Detection of landmines and UXO using advanced synthetic aperture radar technology," Proceedings of Spie, 2016, pp. 1–8.

[18]L. Moreira et al., "A Drone-borne Multiband DInSAR: Results and Applications," 2019 IEEE Radar Conference, MA, USA, Apr. 2019, p. 1-6.

[19]K. Oshea, R. Nash, "An Introduction to Convolutional Neural Networks," Neural and Evolutionary Computing, 2015, pp. 1–11.

[20]J. Hidefonso, E. Aguiar, R. Mauri, E. Barsanulfo, "Susceptibility of Five Forest Species to Coptotermes Gestroi," R. Árvore, Viçosa, vol. 33, no. 6, 2009, pp. 1043–1050.

[21]B. Denardin, E. Pagel, N. Yoshihiro, R. Tuyoshi, N. Rosot, C. Gracioli, "Relação da Morfometria e Competição com o Crescimento de Trichilia claussenii em um Fragmento de Floresta Semidecidual, RS," Floresta, Curitiba, PR, vol. 45, no. 2, 2015, pp. 373 – 382.

[22]A. Moreira, L. Forti, A. Andrade, M. Boaretto, J. Lopes, "Nest Architecture of Atta laevigata (F. Smith, 1858) (Hymenoptera: Formicidae)," Studies on Neotropical Fauna and Environment, vol. 39, no. 2, 2004, pp. 109–116.

[23]L. Grandeza, J. Moraes, R. Zanetti, "Estimativa do Crescimento Externo de Ninhos de Atta sexdens rubropilosa e Atta laevigata (F. Smith) (Hymenoptera: Formicidae) em Áreas de Reflorestamento com Eucalipto," An. Soc. Entomol. vol. 28, no. 1, 1999, pp. 59–64.

[24]Computer Simulation Technology (CST) Studio Suite 2022. https://www.3ds.com/products-services/simulia/products/cst-studio-suite.

[25]C. Parizotto, E. Oglio, L. Vasconcelos, P. Sousa Jr. E. Filho, C. Kuhnen, "Measuring Dielectric Properties for Microwave-assisted Extraction of Essential Oils Using Single-mode and Multimode Reactors," Royal Society of Chemistry, vol. 9, 2019, pp. 5259–5269.

[26]Tree Functional Attributes and Ecological Database, "Wood Density of Species for Genus: Eucalyptus," http://db.worldagroforestry.org/ wd/species/Eucalyptus, (webpage accessed on Apr. 10th, 2023).

[27]D. Pretto, "Arquitetura dos Tuneis de Forrageamento e do Ninho de Atta Sexdens Rubropilosa Forel, 1908 (Hymenoptera-Formicidae), Dispersão de Substrato e Dinamica do Inseticida na Colonia," MSc. Thesis, Faculdade de Ciencias Agronomicas, UNESP, SP, Brazil, 1996.

[28]J. Goes, V. Castro, L. Bins, H. Hernandez-Figueroa, "Spiral SAR Imaging with Fast Factorized Back-Projection: A Phase Error Analysis," Sensors, vol. 21, no. 15, 2021, pp. 1–20.

[29]A. John, "Koppe Climate Classification," Encyclopedia Britannica, https://www.britannica.com/science/Koppen-climate-classification, (webpage accessed on Apr. 10th, 2023).

[30]C. Alvares, J. Stape, P. Sentelhas, J. Gonçalves, G. Sparovek, "Koppen`s Climate Classification Map for Brazil," Meteorologische Zeitschrift, vol. 22, no. 6, 2014, pp. 711–728.

[31]D. Feng, D. An, L. Chen, X. Huang, and Z. Zhou, "Multicircular SAR 3-D Imaging Based on Iterative Adaptive Approach," in 2019 6th APSAR, Nov. 2019, pp. 1–5.

[32]R. Zanetti, J. Zanuncio, J. Santos, W. Da Silva, G. Ribeiro, P. Lemes, "An Overview of Integrated Management of Leaf-Cutting Ants (Hymenoptera: Formicidae) in Brazilian Forest Plantations," Forests, vol. 5, 2014, pp. 439–454.

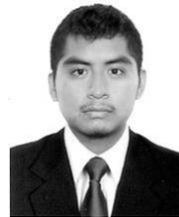

**Gian Oré**, Gian Oré a BSc degree in Electronic Engineering from Universidad Peruana de Ciencias Aplicadas (UPC), Lima, Perú, in 2016, and an MSc degree in Electrical Engineering from the University of Campinas (UNICAMP), Campinas, Sao Paulo, Brazil in 2021. He is currently pursuing a PhD degree in Electrical Engineering at UNICAMP. His research interests include remote sensing, interferometry SAR, signal processing, and deep Learning for SAR images.

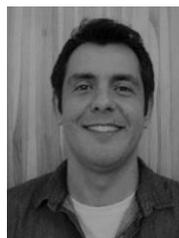

**Alexandre dos Santos**, Forest Engineering and MSc in Entomology from the Federal University of Lavras (UFLA), Lavras, Minas Gerais State, Brazil, Dr. in Entomology from the Federal University of Lavras/Unit Biostatistique et Processus Spatiaux (BioSP) (INRA/France). He is a Professor of Forest Engineering at IFMT, Cáceres, Mato Grosso State, Brazil, and a Postgraduate Program in Entomology Professor at UFLA, Lavras, Minas Gerais state, Brazil.

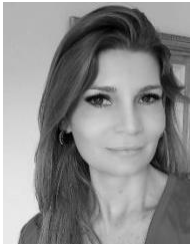

**Daniele Ukan** is a forest engineer with a PhD and a professor at the Department of Forest Engineering at Midwestern State University in Parana, Brazil.

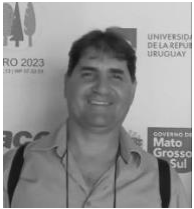

**Ronald Zanetti**, Full Professor at the Department of Entomology, Federal University of Lavras (UFLA), Minas Gerais, Brazil. Forest Engineer (UFV-1990), MSc in Entomology (UFV-1992), PhD in Forestry Sciences (UFV-1998), and Postdoc at Lancaster University (UK). Professor at the Graduate Programs in Entomology and Applied Ecology at UFLA. Coordinator (2016-2021) and Vice-coordinator (2008-2012) of the Graduate Program in Entomology at UFLA.

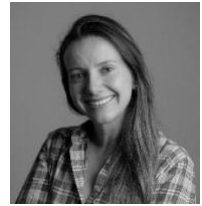

**Mariane Camargo** is a Forestry Engineer with an MSc in forestry sciences from the Midwest State University, Parana, Brazil, and she is a PhD student at the same institution. She is a specialist researcher at Klabin S/A, emphasizing forest protection, silviculture, and water conservation management.

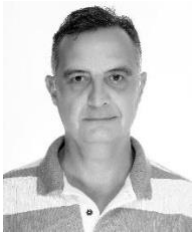

**Luciano P. Oliveira**, PhD in Electrical Engineering from the University of Campinas (UNICAMP), Campinas, Sao Paulo, Brazil, in 2013. He is a Lead Researcher at the Direct Energy Research Center at the Technology Innovation Institute (TII) in Abu Dhabi, UAE, and a Collaborating Researcher at the School of Electrical and Computer Engineering at UNICAMP.

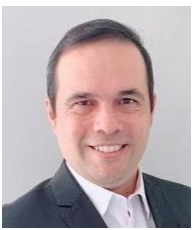

**Guillermo Kemper**, an electronic engineer who graduated from Universidad Privada Antenor Orrego de Trujillo, Peru (1994). He obtained the degrees of PhD (2001) and Master (1996) in Electronic Engineering in the areas of Telecommunications and Telematics from the University of Campinas (UNICAMP), Campinas, Sao Paulo, Brazil. His line of research focuses on digital signal and image processing, emphasizing digital television, biomedical engineering, teledetection, and precision agriculture, among others. He is currently a full-time research professor at the Faculty of Engineering - School of Electronic Engineering of the Universidad Peruana de Ciencias Aplicadas (UPC), Lima, Peru, and a senior member of the IEEE.

**Alonso Sanchez** received his MSc in Electronics Engineering from Universidad Nacional de Ingeniería (UNI) and his BSc in Electronics Engineering from Universidad Peruana de Ciencias

Aplicadas (UPC), both in Lima, Peru. He is currently an undergraduate professor at the Electronics Engineering Programme at UPC and a postgraduate professor at the Faculty of Electrical and Electronics Engineering at UNI (FIEE-UNI). His main research interests are multimedia coding, medical image processing, machine learning, and deep learning.

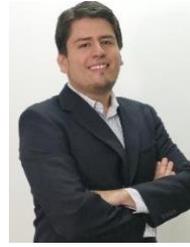

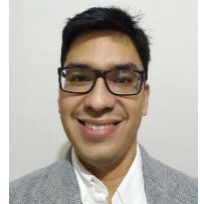

**Aldo André Díaz-Salazar** is an Assistant Professor of Artificial Intelligence for Robotics and Autonomous Systems at the Federal University of Goiás, Goiania, Brazil. His research interests include CDIO and STEM education, robotics, computer vision, sensor fusion, artificial intelligence, and signal processing. His projects involve estimating motion from cameras and heterogeneous sensors equipped on mobile platforms carried by robots and users.

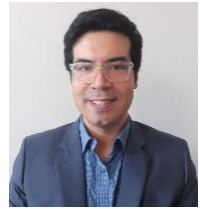

**Jorge Gonzalez** is an assistant professor at the Computer Science Department of the University of Technology and Engineering (UTEC), Lima, Peru. He obtained his PhD from the University of Campinas (UNICAMP), Campinas, Sao Paulo, Brazil. His current research interests are in computer architecture, optical interconnects, memory systems, and intra-chip traffic.

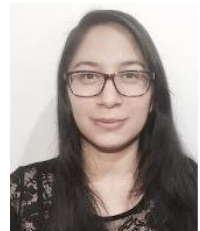

**Ruth Rubio-Noriega** is the Telecommunications and Information Technologies group leader at the National Institute for Research and Training in Telecommunications (INICTEL) and is a lecturer at the National University of Engineering (UNI) in Lima, Peru. She has a PhD and MSc in Telecommunications and Telematics from the University of Campinas (UNICAMP), Campinas, Sao Paulo, Brazil, in 2017. Her research areas include integrated photonics, optoelectronic devices, optical interconnects, microwaves for optical devices, and applied computational electromagnetism.

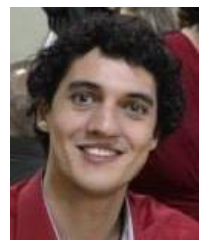

**Levy Boccato** received his BSc (2008) in Computer Engineering, MSc (2010), and PhD (2013) in Electrical Engineering, all from the University of Campinas (UNICAMP), Campinas, Sao Paulo, Brazil. Currently, he is an Assistant Professor at the same university. His research interests include computational intelligence, signal processing, adaptive filtering, brain-computer interfaces, and machine learning.

**Hugo E. Hernandez-Figueroa**, PhD in physics from the Imperial College London, in 1994. Full Professor at the

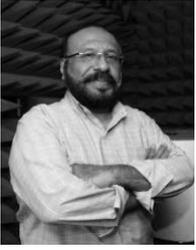 University of Campinas (UNICAMP), School of Electrical and Computer Engineering (FEEC), Campinas, Sao Paulo, Brazil, since 2005. Dean of FEEC (2023-2027). Fellow of OPTICA (2011) and a recipient of the IEEE Third Millennium Medal in 2000. Advisory Committee Member in Engineering at the Sao Paulo State Science Foundation (FAPESP) since 2014. His research interests include integrated photonics, radars, photonics biosensors, and metamaterials for communications and biotechnology.